# A "skewed" lognormal approximation to the probability distribution function of the large-scale density field

S. Colombi

NASA/Fermilab Astrophysics Center, Fermi National Accelerator Laboratory,
P.O. Box 500, Batavia, IL 60510

**Abstract**

I propose a method to fit the probability distribution function (hereafter PDF) of the large scale density field $\rho$, motivated by a Lagrangian version of the continuity equation. It consists in applying the Edgeworth expansion to the quantity $\Phi \equiv \log \rho - \langle \log \rho \rangle$. The method is tested on the matter particle distribution in two cold dark matter $N$-body simulations of different physical sizes to cover a large dynamic range. It is seen to be very efficient, even in the non-linear regime, and may thus be used as an analytical tool to study the effect on the PDF of the transition between the weakly non-linear regime and the highly non-linear regime.

**Subject headings**: cosmology: large-scale structure of universe – galaxies: clustering

# 1 Introduction

It is generally believed that the large scale structures of the observed galaxy distribution have grown under gravitational instability from small initial fluctuations. Many statistical tools exist to compare models to the observations. I consider here the probability distribution function $P(\rho, \ell)$ (PDF) of the density field smoothed with a filter window of typical size $\ell$, which can be a spherical (cubical) cell of radius (size) $\ell$ or a gaussian of half-width $\ell$. Possible discreteness effects will be neglected and the initial fluctuations will be supposed gaussian.

In the weakly non-linear regime, e.g., in the case the fluctuations are small, one can use perturbation theory to analytically predict the behaviour of the PDF (Bernardeau 1992, Kofman et al. 1993, hereafter KBGND, Juszkiewicz et al. 1993, hereafter JWACB). For example, JWACB have suggested to use the Edgeworth expansion to fit the shape of the PDF in the weakly non-linear regime and to evaluate how it deviates from the gaussian limit. Given the variance $\sigma^2(\ell) = \langle \delta^2 \rangle$ of the smoothed density contrast $\delta \equiv \rho/\langle \rho \rangle - 1$, this method consists in expanding the function $P(\rho, \ell)$ as a series of terms in powers of $\sigma$ relatively to a gaussian of same variance and same average. But this method is practically valid only for a small or moderate variance $\sigma^2 \lesssim 1$.

The non-linear case is more complicated and numerical simulations are generally used to study it. The PDF, as measured in $N$-body simulations as a function of $\rho$ in the non-linear regime, is strongly non gaussian. It is seen to present a power-law behaviour surrounded by two exponential cut-offs (Bouchet, Schaeffer & Davis 1991, Bouchet & Hernquist 1993). Such a behaviour was predicted by Balian & Schaeffer (1988, 1989, see also Fry 1985), but their calculation needs supplementary assumptions on the correlation hierarchy, that are not proved to be necessarily valid.

Fortunately, the continuity equation can be used as a filter to the non-linear regime. We can write the density field as

$$\rho = \langle \rho \rangle e^{\overline{\Phi}}, \quad \overline{\Phi} = -\int_{\text{trajectory}} \nabla_{\mathbf{x}} \cdot \mathbf{v} \frac{dt}{a}. \tag{1}$$

In equation (1), $a$ is the expansion factor, $\mathbf{x}$ is the comoving coordinate and $\mathbf{v}$ the peculiar velocity $[\mathbf{v}/a = d\mathbf{x}/dt]$. The integral is done in a Lagrangian way, e.g., by following the particle trajectories. Such an expression is thus valid only before shell crossing. In the the weakly non-linear regime and if standard linear theory applies, $\overline{\Phi}$ is proportional to $\nabla_{\mathbf{x}} \cdot \mathbf{v}$ and is thus gaussian. If, instead of expanding it at first order, we keep the exponential of equation (1) to insure the positivity of the density, we see that $\rho$ is rather lognormal than gaussian, as already pointed out by Coles & Jones (1991).

The validity of this simple reasoning is confirmed by recent measurements on the observed three dimensional galaxy catalogs (Hamilton 1985, Bouchet et al. 1993) as well as on cold dark matter $N$-body simulations (KBGND), that indicate that the lognormal distribution (hereafter LNDF) is a very good fit to the data. Note that Hubble already used the LNDF in 1934 to fit the PDF measured on the projected galaxy distribution.

This suggests we apply the Edgeworth expansion to the variable $\overline{\Phi}$ rather than the density itself and use equation (1) as a non-linear filter to insure the positivity of $\rho$. This letter is thus organized as follows. In Sect. 2, I recall the Edgeworth expansion (up to third order) and compute a fit to the PDF of the smoothed density field, applying this approximation to the measured variable $\overline{\Phi}$, or more exactly, to the measured variable $\Phi \equiv \log \rho - \langle \log \rho \rangle$ where $\rho$ is the smoothed density field. I will relate the moments of the variable $\rho$ to those of the variable $\Phi$. In Sect. 3, I test the performance of the approximation by measuring the PDF on the distribution of matter particles from two cold dark matter (hereafter CDM) $N$-body simulations. Section 4 is the conclusion.

## 2 The "skewed" lognormal approximation

In the following, I will not exactly consider the quantity $\overline{\Phi}$ given by equation (1), but, for the sake of simplicity, the quantity of zero average $\Phi \equiv \log \rho - \langle \log \rho \rangle$, where $\rho$ is the smoothed density field. I will also remove the scale dependence of the PDF, although it is implicitly assumed.

Let $\sigma_\Phi^2 = \langle \Phi^2 \rangle$ be the variance of $\Phi$. At third order in $\sigma_\Phi$, the Edgeworth approximation is written (e.g., JWACB)

$$P_{E,3}(\Phi)d\Phi = \left[1 + \frac{1}{3!}T_3\sigma_\Phi H_3(\nu) + \frac{1}{4!}T_4\sigma_\Phi^2 H_4(\nu) + \frac{10}{6!}T_3^2\sigma_\Phi^2 H_6(\nu)\right]G(\nu)d\nu, \tag{2}$$

where $\nu \equiv \Phi/\sigma_\Phi$, $H_m(x) \equiv d^m[\exp(-x^2/2)]/dx^m$ is the Hermite polynomial of degree $m$, $G(\nu)$ a gaussian of average zero and variance unity. The quantities $T_3(\ell)$ and $T_4(\ell)$ are the renormalized skewness and the renormalized kurtosis of the field $\Phi$:

$$T_3 \equiv \frac{\langle \Phi^3 \rangle}{\sigma_\Phi^4}, \quad T_4 \equiv \frac{\langle \Phi^4 \rangle - 3\sigma_\Phi^4}{\sigma_\Phi^6}. \tag{3}$$

The average $\langle \Phi^Q \rangle$ is defined by $\langle \Phi^Q \rangle \equiv \int \Phi^Q P(\Phi) d\Phi$. The writing (2) insures that the variance, the skewness and the kurtosis of the PDF $P_{E,3}(\Phi)$ are also $\sigma_\Phi^2$, $T_3$ and $T_4$. At first order in $\sigma_\Phi$, the Edgeworth approximation $P_{E,1}$ would simply give a gaussian. The second order correction in $\sigma_\Phi$ comes down to taking into account the fact that the PDF is asymmetric, and thus a term proportional to $T_3\sigma_\Phi H_3(\nu)$ appears in $P_{E,2}$. The third order takes into account the flattening of the PDF. We can see that the function $P_{E,k}(\Phi)$ is not really a distribution function, since it is not generally positive definite and thus has to be considered only as an approximation valid at $k^{\text{th}}$ order in $\sigma$.

The PDF of the density field can now be approximated by the function

$$P_k(\delta)d\delta = P_k(\rho)d\rho = \frac{1}{\rho}P_{E,k}[\log(\rho/\overline{\rho}_k)]d\rho, \tag{4}$$

where the factor

$$\overline{\rho}_k = \frac{\langle \rho \rangle}{\int P_{E,k}(\log x)dx} \tag{5}$$

insures mass conservation, e.g., $\langle \rho \rangle_k \equiv \int \rho P_k(\rho)d\rho = \langle \rho \rangle$.

The function $P_1(\rho) = (\sqrt{2\pi}\rho\sigma_\Phi)^{-1} \exp\{-[\log(\rho/\overline{\rho}_1)]^2/\sigma_\Phi^2\}$ is nothing but a LNDF. Since the function $P_k(\rho)$ [or equivalently $P_k(\delta)$] involves some corrections to the LNDF when $k \geq 2$, I will call it a "skewed" LNDF, hereafter SLNDF$k$, $k = 1, 2, 3$.

The moment of order $Q$ of the SLNDF$k$ can be analytically evaluated. Let us write it as $\langle \rho^Q \rangle_k \equiv \int \rho^Q P_k(\rho)d\rho$. We easily have

$$\langle \rho^Q \rangle_k = \langle \rho \rangle^Q f_k(Q)[f_k(1)]^{-Q} \exp\left[Q(Q-1)\sigma_\Phi^2/2\right], \tag{6}$$

with

$$f_1(Q) = 1, \tag{7}$$

$$f_2(Q) = 1 + \frac{1}{6}Q^3 T_3 \sigma_\Phi^4, \tag{8}$$

$$f_3(Q) = 1 + \frac{1}{6}Q^3 T_3 \sigma_\Phi^4 + \frac{1}{24}Q^4 T_4 \sigma_\Phi^6 + \frac{1}{72}Q^6 T_3^2 \sigma_\Phi^8. \tag{9}$$

Table 1: Values of statistical estimators in various approximations for the CDM case

| $\ell(\mathrm{Mpc})$ | $\hat{\sigma}^2$ | $\hat{\sigma}_\Phi^2$ | $\hat{\sigma}_\Phi \hat{T}_3$ | $\hat{\sigma}_\Phi^2 \hat{T}_4$ | $\hat{S}_3$ | $\hat{S}_3^{\mathrm{cor}}$ | $S_3^{\mathrm{wea}}$ | $\tilde{S}_{3,1}$ | $\tilde{S}_{3,2}$ | $\tilde{S}_{3,3}$ |
|---|---|---|---|---|---|---|---|---|---|---|
| 2.5 | 9.3 | 1.4 | 0.55 | 0.45 | 5.0 | 7.2 | – | 6.0 | 6.1 | 7.5 |
| 4.0 | 3.6 | 1.2 | 0.40 | 0.25 | 4.2 | 6.3 | – | 5.2 | 5.7 | 7.4 |
| 6.4 | 1.5 | 0.8 | 0.25 | -0.06 | 3.6 | 5.1 | – | 4.2 | 4.9 | 5.6 |
| 12 | 0.6 | 0.6 | -0.15 | 0.06 | 4.2 | 4.2 | 3.4 | 3.8 | 2.8 | 3.6 |
| 23 | 0.2 | 0.2 | -0.06 | -0.03 | 3.2 | 3.2 | 3.0 | 3.2 | 3.0 | 2.9 |
| 46 | 0.04 | 0.04 | -0.20 | -0.15 | 2.0 | n.a. | 2.6 | 3.0 | 2.1 | 2.0 |

Let $\sigma^2 \equiv \langle \delta^2 \rangle$ be the variance of the density contrast, and $S_3$ and $S_4$ the skewness and kurtosis of $\delta$, defined in a similar way as $T_3$ and $T_4$ (e.g., eq. [3]). Let $\tilde{\sigma}_k^2 = \langle \delta^2 \rangle_k$, $\tilde{S}_{3,k}$ and $\tilde{S}_{4,k}$ be the variance, the skewness and the kurtosis of the SLNDF$k$. In general, we will have $\sigma^2 \neq \tilde{\sigma}_k^2$, $S_i \neq \tilde{S}_{i,k}$, $i = 3, 4$ except in the weakly non-linear limit (but it depends on the quantity considered and on $k$, see Appendix). Only the averages of $P_k(\rho)$ and $P(\rho)$ are equal. Note that the PDF $P_k(\rho)$ is not used in the usual way to fit the PDF $P(\rho)$, e.g., by insuring that the low-order moments of the density distribution are equal for both the fit and the real PDF. It is rather used in such a way that the low-order moments of the field $\Phi$ are equal for both the approximation $P_{E,k}(\Phi)$ and the real PDF $P(\Phi)$.

The exponential function in equation (6) makes $\sigma^2$ generally much larger than $\sigma_\Phi^2$ in the non-linear regime. For example, in the lognormal case ($k = 1$), we have $\sigma^2 = \exp(\sigma_\Phi^2)$ and hence $\sigma^2 \gtrsim 2.7$ when $\sigma_\Phi^2 \gtrsim 1$. The Edgeworth expansion was seen by JWACB to be practically valid for $\sigma_\Phi^2 \lesssim 1$. We can thus believe that the SLDNF$k$ should be a good fit of $P(\rho)$ in the non-linear regime, at least for moderate values of $\sigma^2$, of order a few unities.

## 3 Application: measurement of the PDF in CDM samples

Bouchet et al. (1991) and Colombi et al. (1994, hereafter CBS) have measured the PDF for cubical cells of size $\ell$ in the distribution of matter particles coming from two $N$-body CDM simulations generated with a P$^3$M code and involving 262 144 matter particles. The first simulation (Davis & Efstathiou 1988), hereafter CDM1, has a moderate physical size $L_1 = 64$ Mpc (with $H_0 = 50$ km/s/Mpc), which permits us to measure the PDF in the non-linear regime ($\langle \delta^2 \rangle \gtrsim 1$). The second simulation (Frenk et al. 1990), hereafter CDM2, has a large size $L_2 = 360$ Mpc and is thus appropriate to measure the PDF in the weakly non-linear regime ($\langle \delta^2 \rangle \lesssim 1$). The two simulations were both stopped when the variance of the distribution in a sphere of radius 16 Mpc was $1/b^2$ with $b \sim 2.4$ so they are two different realizations of the same underlying statistics. The first column of table 1 gives the scale in Mpc at which the PDF has been measured.

Since the samples considered here are sets of points, their discrete nature has to be taken into account. Let $\hat{P}(\hat{\rho})$ be the measured PDF and $P(\rho)$ the real PDF [e.g., the continuum limit of $\hat{P}(\hat{\rho})$]. Within a normalization factor, $\hat{P}(\hat{\rho})$ is nothing but the count probability $P_N(\ell)$ (hereafter CPDF), defined as the probability of having $N$ objects in a cell of size $\ell$ thrown at random in the sample. To estimate $\sigma_\Phi^2$, $T_3$ and $T_4$, I use the indicator $\hat{\Phi} \equiv \log \hat{\rho} - \langle \log \hat{\rho} \rangle$, with $\hat{\rho} \equiv N/\langle N \rangle$ and thus

$$\langle \log \hat{\rho} \rangle \equiv \sum_{N \geq 1} \log(N/\langle N \rangle) P_N, \tag{10}$$

where $\langle N \rangle = \sum N P_N$. The moment of order $Q$ of $\hat{\Phi}$ will be estimated by

$$\langle \hat{\Phi}^Q \rangle = \sum_{N \geq 1} [\log(N/\langle N \rangle) - \langle \log \hat{\rho} \rangle]^Q P_N. \tag{11}$$

But the sums (10) and (11) give a lot of weight to the small values of $N$, that are contaminated by discreteness effects and may thus provide a very bad estimation of the real moment of order $Q$ of the variable $\Phi$. However, if the measured function $P_N(\ell)$ presents an exponential cut-off at small $N$, the contamination by discreteness will be small. Such a cut-off is expected if the cell size is large compared to the mean distance between two objects in an underdense region. Practically, I simply imposed the maximum of the CPDF to be at $N > 1$. But since the discrete nature of the CPDF is intrinsically taken into account in the sums (10) and (11), the function really fitted will be the measured CPDF and not the PDF of the underlying continuous density field.

The second and third column of Table 1 give respectively the measured quantities $\hat{\sigma}^2$ and $\hat{\sigma}^2_\Phi$. The variance of the field $\hat{\Phi}$, is, as expected (see Appendix), equal to $\hat{\sigma}^2$ in the weakly non-linear limit. In the non-linear regime, $\hat{\sigma}^2_\Phi$ increases in a way much slower than $\hat{\sigma}^2$ and is till only of order unity when $\hat{\sigma}^2 \sim 10$, in agreement with the discussion of the end of previous section.

Figure 1 gives the measured "mass" distribution function $\hat{P}(\hat{M}, \ell)$ as a function of $\hat{M}$, compared with the SLNDF$k$, $k = 1, 2, 3$. To have the same mass scales in CDM1 and CDM2, I take $\hat{M} = (L_i/L_1)^3 N$, where $N$ is the number of objects measured in a cell thrown at random in CDM$i$. The variance of the PDF measured at each scale, increases with $\max[\hat{P}(\hat{M}, \ell)]$ and crosses unity between the third and the fourth curve.

We see that the LNDF ($k = 1$), although a first order approximation, is already a good fit, particularly in the weakly non-linear regime, where it is impossible to distinguish it from the measurement, as noticed earlier by KBGND. But this property is certainly a particular feature of the CDM matter distribution. Indeed, a careful examination of equation (2) shows that the quantities $\sigma_\Phi T_3$ and $\sigma^2_\Phi T_4$ in some way estimate the low-order corrections to the LNDF needed to fit the measured PDF. Equations (13) and (14) of Appendix say then that if, in the weakly non-linear regime, $S_3 \simeq 3$ and $S_4 \simeq 16$, then $T_3$ and $T_4$ should be small and hence $\sigma_\Phi T_3$ and $\sigma^2_\Phi T_4$ even smaller. Their measured values are given in the fourth and third columns of Table 1. The evaluation of $S_3$ in the regime $\sigma^2 \ll 1$ can be done by using second order perturbation theory (e.g., Juszkiewicz et al. 1993). The result is given in the 8$^{\text{th}}$ column of Table 1 [$S_3^{\text{wea}}$] and we indeed have $S_3 \simeq 3$ in this regime.

When $\hat{\sigma}^2 \gtrsim 1$, the LNDF does not fit very well the global shape of the measured CPDF, but the second order approximation SLNDF2 is much better and the third order one SLNDF3 is almost a perfect fit (if one forgets the low $M$ part of the curves, that is anyway contaminated by discreteness effects). The reason for that is that the corrective terms $\hat{\sigma}_\Phi \hat{T}_3$ and $\hat{\sigma}^2_\Phi \hat{T}_4$ are not, in this regime, a small fraction of unity. We can however say that the PDF measured in the CDM matter distribution does not deviate "very much" from a LNDF because we still have $|\hat{\sigma}^2_\phi \hat{T}_4| \lesssim |\hat{\sigma}_\Phi \hat{T}_3| \lesssim 1$. This last inequality also explains why the SLDNF$k$ is a good approximation, since the amount of the correction seems to decrease with increasing $k$.

The last six columns of Table 1. give the measured values of the skewness in various approximations. The 6$^{\text{th}}$ column gives the direct measurement of $S_3$ in CDM1 and CDM2, the next one gives the more realistic values $S_3^{\text{cor}}$ of the skewness obtained when correcting the measured CPDF for finite volume effects by using the method explained in CBS. The three last columns give the measured $\tilde{S}_{3,k}$, $k = 1, 2, 3$. The LNDF is only a first order approximation ($k = 1$) and should not give the appropriate value of $S_3$, even in the weakly non-linear regime (see Appendix). But in the case of the CDM model, it provides as discussed above a reasonable approximation of $S_3$. The measured $\tilde{S}_{3,2}$ and $\tilde{S}_{3,3}$ give, as expected, a rather good estimate of $S_3$ in the weakly non-linear regime. In the non-linear regime, $\tilde{S}_{3,2}$ tends to underestimate the value of $S_3$ derived from the measured CPDF, whereas $\tilde{S}_{3,3}$ tends to overestimate it. Note that the measured numbers $\tilde{S}_{3,k}$, $k=2,3$ are much closer to the real value of $S_3$ (corrected for finite volume effects, e.g., $S_3^{\rm cor}$) than to the direct measurement (e.g., $\hat{S}_3$). Therefore, when both used, they provide (in this particular example) good estimators of the skewness.

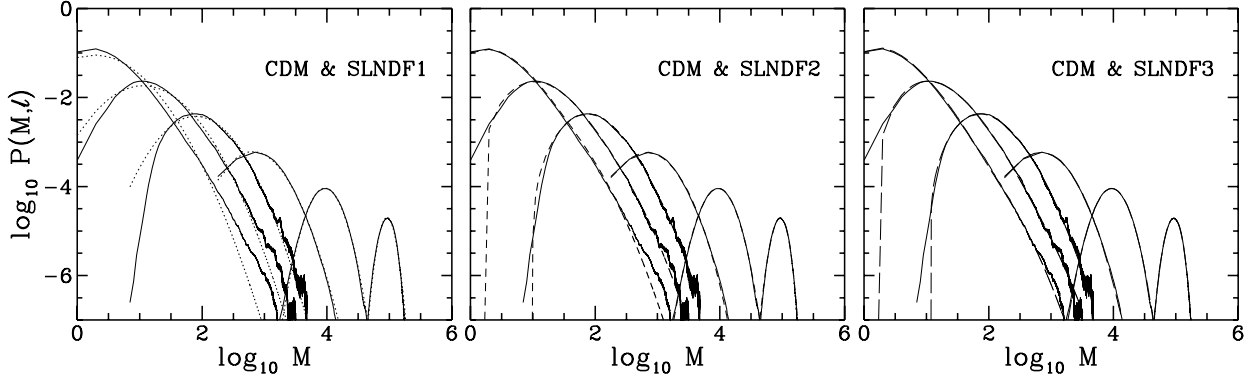

Figure 1: The PDF as measured in the two samples CDM1 and CDM2 (solid curves) compared to the "skewed" lognormal approximation SLNDF$k$, $k=1$ (dots in left panel), $k=2$ (shord dashes in middle panel) and $k=3$ (long dashes in right panel). In each panel, the three lefts curves correspond to the measurement on CDM1 and the three right curves to the measurement on CDM2. The quantity represented is the "mass" PDF $\hat{P}(\hat{M},\ell)$ as a function of $\hat{M}$ for various scales (see text and Table 1) in logarithmic coordinates. The fit is given only for the available values of $\hat{M}$ (e.g., for which the measured $\hat{P}(\hat{M},\ell)$ is not vanishing), which explains the possible interruption of the curves at low $\hat{M}$ and at high $\hat{M}$. The Edgeworth expansion, when directly applied to the PDF, would provide a good fit (improving with increasing order $k$) only in the weakly non-linear regime.

## 4 Conclusion

In this letter, I have proposed a new approximation to fit the probability distribution function (PDF) of the large scale density field $\rho$, which I call the "skewed" lognormal approximation (SLNDF). It consists in applying the Edgeworth expansion (e.g., JWACB) to the PDF of the statistical quantity $\Phi \equiv \log\rho - \langle\log\rho\rangle$. This idea is motivated by writing the continuity equation in a Lagrangian way. The SLNDF has been tested on the matter distribution of particles from two cold dark matter (CDM) simulations, one of large physical size and one of small physical size to cover a large dynamical range. It is seen through this example to provide a very good fit to the measured PDF, even in the non-linear regime. This approximation should not only be valid in the particular case of the matter distribution in a CDM universe. A more general study is in progress, concerning the measurement of the CPDF in scale invariant simulations $[\langle|\delta_k|^2\rangle \propto k^n$, with $n = 1, 0, -1, -2]$ and the preliminary results indeed confirm the efficiency of the approximation (Colombi, Bouchet & Hernquist 1994).

The SLNDF is not positive definite, so it is not a real PDF and it has to be used with caution. However, when used at the appropriate order, it provides a new estimator of the skewness $S_3$ of the PDF. It can be used to study the behaviour of the PDF in both the weakly non-linear regime and the non-linear regime. It may thus permit a better understanding of the transition toward the non-linear regime.

SC thanks F.R. Bouchet, J.A. Frieman and A. Stebbins for useful comments. SC is supported by DOE and by NASA through grant NAGW-2381 at Fermilab. Part of this work was done while SC was at the Institut d'Astrophysique de Paris (CNRS), supported by Ecole Polytechnique.

## Appendix : values of the variance, the skewness and the kurtosis of the SLNDF$k$ in the weakly non-linear regime

In the weakly non-linear limit $\sigma^2 \ll 1$, we have, with the notation of Sect. 2,

$$\tilde{\sigma}_1^2 = \sigma^2 + \mathcal{O}^4 = \sigma_\Phi^2 + \mathcal{O}^4, \quad \tilde{S}_{3,1} = 3 + \mathcal{O}^2, \quad \tilde{S}_{4,1} = 16 + \mathcal{O}^2, \tag{12}$$

$$\begin{aligned}
\tilde{\sigma}_2^2 &= \sigma^2 + \mathcal{O}^6 = \sigma_\phi^2 + (\tfrac{1}{2} + T_3)\sigma_\phi^4 + \mathcal{O}^6, \\
\tilde{S}_{3,2} &= S_3 + \mathcal{O}^2 = 3 + T_3 + \mathcal{O}^2, \quad \tilde{S}_{4,2} = 16 + 12T_3 + \mathcal{O}^2,
\end{aligned} \tag{13}$$

$$\begin{aligned}
\tilde{\sigma}_3^2 &= \sigma^2 + \mathcal{O}^8 = \sigma_\phi^2 + (\tfrac{1}{2} + T_3)\sigma_\phi^4 + (\tfrac{1}{6} + T_3 + \tfrac{7}{12}T_4)\sigma_\phi^6 + \mathcal{O}^8, \\
\tilde{S}_{3,3} &= S_3 + \mathcal{O}^4 = 3 + T_3 + (1 + 2T_3 - 2T_3^2 + \tfrac{3}{2}T_4)\sigma_\phi^2 + \mathcal{O}^4, \\
\tilde{S}_{4,3} &= S_4 + \mathcal{O}^2 = 16 + 12T_3 + T_4 + \mathcal{O}^2,
\end{aligned} \tag{14}$$

with $\mathcal{O}^n = \mathcal{O}(\sigma^n)$. Thus, in the weakly non-linear regime, the SLNDF1 has the same variance than the PDF, but its skewness and its kurtosis are fixed, $\tilde{S}_{3,1} = 3$, $\tilde{S}_{4,1} = 16$. On the contrary, as far as the real PDF is concerned, $S_3$ and $S_4$ should depend on initial conditions (e.g. Juszkiewicz, Bouchet & Colombi 1993, Bernardeau 1994 and references herein). To have the appropriate skewness, the second-order correction is needed, which fixes $T_3$. The third order approximation only would give the true kurtosis in the weakly non-linear regime, which fixes $T_4$. Note that the SLNDF3 could provide a way to understand the transition toward the non-linear regime on the skewness through second line of equation (14).

## References

Balian, R., & Schaeffer, R. 1988, ApJ, 335, L43
Balian, R., & Schaeffer, R. 1989, A&A, 220, 1
Bernardeau, F. 1991, ApJ, 392, 1
Bernardeau, F. 1993, CITA preprint (CITA/93/44)
Bouchet, F.R., & Hernquist, L. 1992, ApJ, 400, 25
Bouchet, F.R., Schaeffer, R., & Davis, M. 1991, ApJ, 383, 19
Bouchet, F.R., Strauss, M.A., Davis, M., Fisher, K.B., Yahil, A., & Huchra, J.P. 1993, ApJ, 417, 36
Coles, P., & Jones, B. 1991, MNRAS, 248, 1
Colombi, S., Bouchet, F.R., & Schaeffer, R. 1994, A&A, 281, 301 (CBS)
Colombi, S., Bouchet, F.R., & Hernquist, L. 1994, in preparation
Davis, M., & Efstathiou, G. 1988, in Large Scale Motions in the Universe, Proceedings of the Vatican Study Week 1987, Ed. V.C. Rubin & G.V. Coyne, Princeton, Princeton Univ. Press, p. 256
Frenk, C.S., White, S.D.M., Efstathiou, G., & Davis, M. 1990, ApJ, 351, 10

Fry, J.N. 1985, ApJ, 289, 10

Hamilton, A.J.S. 1985, ApJ, 292, L35

Hubble, E. 1934, ApJ, 79, 8

Juszkiewicz, R., Bouchet, F.R., & Colombi, S. 1993, ApJ, 412, L9

Juszkiewicz, R., Weinberg, D.H., Amsterdamski, P., Chodorowski, M., & Bouchet, F.R. 1993, IAS preprint (IASSNS-AST 93/50) (JWACB)

Kofman, L., Bertschinger, E., Gelb, J.M., Nusser, A., & Dekel, A. 1994, ApJ, 420, 44 (KBGND)